\documentclass[%
 reprint,
 onecolumn,
 amsmath,amssymb,
 aps,
]{revtex4-2}

\usepackage{graphicx}
\usepackage{dcolumn}
\usepackage{bm}
\usepackage{hyperref}

\begin{document}


\title{Deep learning regression for inverse quantum scattering}

\author{A. C. Maioli}
 \email{alanmaioli90@gmail.com}
\affiliation{Instituto de Ci\^encias Exatas, Universidade Federal Fluminense,\\ 
27213-415 Volta Redonda --- RJ, Brazil}%
\affiliation{Programa de P\'os Gradua\c{c}\~ao em F\'isica Instituto de F\'\i sica, Universidade Federal Fluminense,\\
24210-346 Niter\'oi --- RJ, Brazil.}

\date{\today}

\begin{abstract}
In this work we study the inverse quantum scattering via deep learning regression, which is implemented via a Multilayer Perceptron. A step-by-step method is provided in order to obtain the potential parameters. A circular boundary-wall potential was chosen to exemplify the method. Detailed discussion about the training is provided. A investigation with noisy data is presented and it is observed that the neural network is useful to predict the potential parameters.

\end{abstract}

\maketitle


\section{Introduction}

Machine Learning is a collection of powerful tools that predicts parameters or classify features based on experimental or synthetic data. A plethora of applications exist, such as the reconstruction of porous media \cite{PhysRevE.96.043309}, feature selection by mutual information \cite{1114861}, percolation and fracture propagation in disordered solids \cite{PhysRevE.102.011001}, the behavior of Ising spin-lattice \cite{PhysRevE.102.013307}, a model for turbulent fluxes that recovers spontaneous zonal flow \cite{PhysRevE.101.061201}, classification of complex features in diffraction images \cite{PhysRevE.99.063309} and much more \cite{Vargas_Hern_ndez_2019, yao, robo}.

Recently, two-dimensional quantum scattering is receiving attention, E. de Prunelé gave a formulation for non-isotropic interactions localized on a circle \cite{DePrunele2018}. Maioli \textit{et al} found analytic solutions for the wavefunction scattered by circular and elliptic billiards \cite{Maioli2018, Maioli2019} and presented a scattering with two-potential formalism \cite{Maioli2019b}. They used a boundary-wall potential introduced by M. G. E. da Luz \textit{et al} \cite{Heller1997}. Which is useful to find analytic solutions for the $T$ matrix, the eigenstates, and the scattering solutions for billiards\cite{Zanetti2008}. Along these lines, the BWM provides a significant way to study quantum scattering and electromagnetic wave propagation for TE or TM modes due to the analogy of both physical phenomena \cite{Zanetti2009}. 
On the other hand, inverse scattering problems have a significant role in applied physics, such as the reconstruction of medium properties \cite{Rizzuti_2017}. In this scenario, G. Ariel and H. Diamant  \cite{PhysRevE.102.022110} showed a method to infer the entropy from the structure factor (which can be obtained by quantum scattering), and T. Tyni numerically investigated the two-dimensional inverse scattering with the aid of Saito's uniqueness theorem \cite{Tyni_2020}. G. Fotopoulos and M. Harju \cite{fotopoulos} study how to retrieve the singularities of an unknown potential using the Born approximation.

The purpose is to provide a simple method that obtains the potential parameters based on the scattering data. This type of inverse problem is extensively frequent in scattering physics. It is designated a regression problem in the machine learning vocabulary. The method consists in choosing a potential that models the physical system, then generating synthetic data to train a neural network. Hence, we select a circular boundary wall potential. This geometry is reasonable because we know the analytic solutions for the eigenstates and the scattered states\cite{Maioli2018}. 

 It is well-known that implementing a neural network to solve a regression problem is considered exceedingly good, and the results improve as one adds more hidden layers. However, it can be computationally exhaustive and hard to converge the network's parameters due to the vanishing gradient problem. Therefore we show how to avoid the last difficulty. The trained neural network can predict the correct results even with noisy input data, and the training set is noiseless.

This paper is organized as follows. In Section \ref{sec: method} we present the method, including how the synthetic data was generated (subsection \ref{subsec: syntheticdata}) and the neural network training (subsection \ref{subsec: trainednetwork}). In section \ref{sec: noise}, it is shown that the trained neural network can predict the correct values for the potential parameters. Finally, we conclude the discussions in section \ref{sec: conclusion}.

\section{The method}\label{sec: method}
The main idea is to provide a fast way to find the potential parameters due to the scattering cross length $l(k)$ obtained for the two-dimensional quantum scattering. The scattering cross length is the two-dimensional analog of the scattering cross-section, the usual formulation can be found at \cite{lapidus1982quantum, Maurone1983, Adhikari1986} and a comparison between 2D and 3D formulas \cite{DePrunele2006}. The method embraces a few simple steps, and some hints follow the example selected throughout this work. The steps are:
\begin{enumerate}
    \item Choose the potential that suits the desired physical system. 
    \item Generate synthetic data that will be the input of the neural network. One can use the scattering cross length and other physical information, such as the particle's mass, Plank's constant, etc.  Therefore, the output is the potential parameters.
    \item Build a neural network. The size of the input will be the number of physical quantities necessary to perform the regression.
    \item Train the neural network with synthetic data.
\end{enumerate}
\subsection{First Step: Boundary wall potential}\label{subsec: potential}
Here we use a circular boundary-wall potential that is defined as a line integral
\begin{equation}
    V(\mathbf{r})=\int_{C}\gamma(s) \; \delta^2(\mathbf{r}-\mathbf{r}(s)) \; ds ,
\end{equation}
where $\gamma(s)$ is the strength function, which we set to be constant $\gamma(s)=\gamma_0$, $C$ is a circle of radius $R$, the $\delta^2$ is the two-dimensional Dirac delta. Writing the potential as a Riemann integral, we have
\begin{equation}
    V(\mathbf{r})=\gamma R \int_{-\pi}^\pi  \frac{\delta(\mathbf{r}-R)}{r}\delta(\theta-s) \; ds ,
\end{equation}
one can see that the parameters $\gamma$ and $R$ uniquely define this type of potential, therefore those are the ones which we need to predict. 
\subsection{Second step: Synthetic data}\label{subsec: syntheticdata}

In this subsection is presented an expression for the scattering cross length $l(k)$. It will be employed to generate the synthetic data. Therefore, it is obtained through the analytic solution of the Lippmann-Schwinger equation outside the circle ($r>R$) \cite{Maioli2018},

\begin{equation}\label{eq:psigrande}
    \psi(\mathbf{r})=J_0(kr)+u_0H_0^{(1)}(kr)+2\sum_{n=1}^\infty i^n\left[J_n(kr)+u_nH_n^{(1)}(kr)\right]\cos\left[n(\theta+(-1)^n\alpha)\right],
\end{equation}

where $J_n$ and $H_n^{(1)}$ are the Bessel and Hankel functions of the first kind of order $n$, respectively, $\alpha$ is the angle between the wave vector $\mathbf{k}$ of the plane wave and the $x-$axis, and
\begin{equation}
    u_n=\frac{2 \pi R \gamma \sigma J_n^2(kR)}{1-2\pi R\gamma \sigma J_n(kR)H_n^{(1)}(kR)},
\end{equation}
where $\sigma=(-i/4)(2m/\hbar^2)$. For the sake of simplicity, we set $\alpha=0$, then using the relation $i^nJ_n(kr)=i^{-n}J_{-n}(kr)$ and $i^nH_n^{(1)}(kr)=i^{-n}H_{-n}^{(1)}(kr)$ one can rewrite the eq. (\ref{eq:psigrande})
\begin{equation}
    \psi(\mathbf{r})=e^{ikx}+\sum_{n=-\infty}^\infty i^n u_n H_n^{(1)}(kr)e^{in\theta},
\end{equation}
where the sum of Bessel functions was identified as the exponential. Along these lines, one can use the asymptotic expansion of the Hankel function
\begin{equation}
    H_n^{(1)}(kr)\approx \sqrt{\frac{2}{\pi k}} \; e^{-i \pi / 4} \; e^{ikr} \; e^{-in\theta /2},
\end{equation}
then it is easy to find the scattering amplitude $f(\theta)$ using
\begin{equation}
    \psi(\mathbf{r}) \approx e^{ikx}+\frac{e^{ikr}}{\sqrt{r}}f(\theta),
\end{equation}
therefore
\begin{equation}\label{eq: amplitudemeu}
    f(\theta)=\sqrt{\frac{2}{\pi k}} \; e^{-i\pi/4} \sum_{n=-\infty}^\infty u_n e^{in\theta}.
\end{equation}

\begin{figure}
    \centering
    \includegraphics[width=.5\columnwidth]{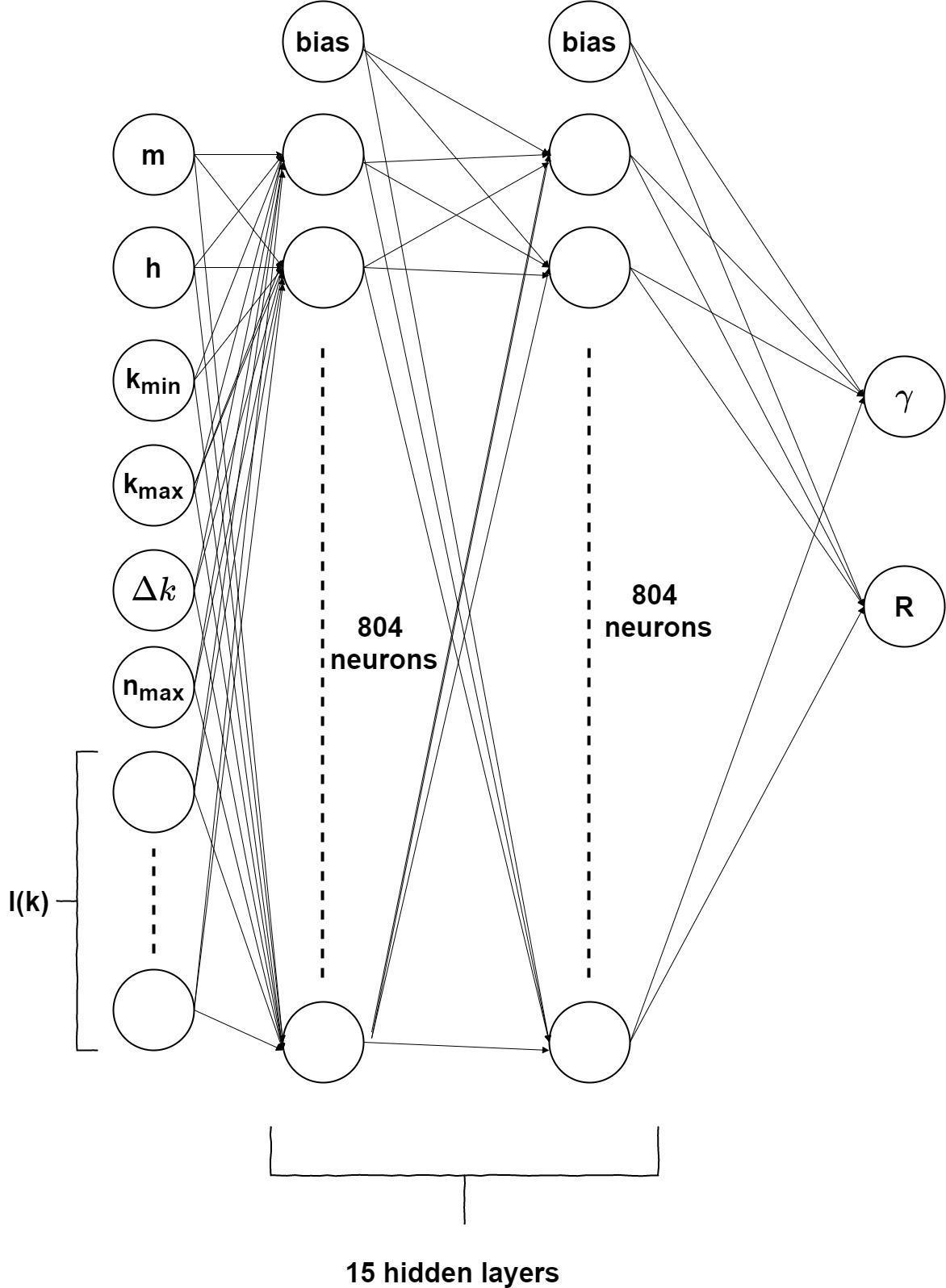}
    \caption{A schematic representation of the neural network. The input have $603$ values, which is defined by $\mathbf{x}=(m,\hbar,k_{min},k_{max},\Delta k,n_{max},l(k_{min}),...,l(k_{max}))^T$. The output contains two values $R$ and $\gamma$. Each hidden layer has $804$ neurons, and there are 15 hidden layers.}
    \label{fig: redeneural}
\end{figure}

\begin{figure}
    \centering
    \includegraphics[width=.7\columnwidth]{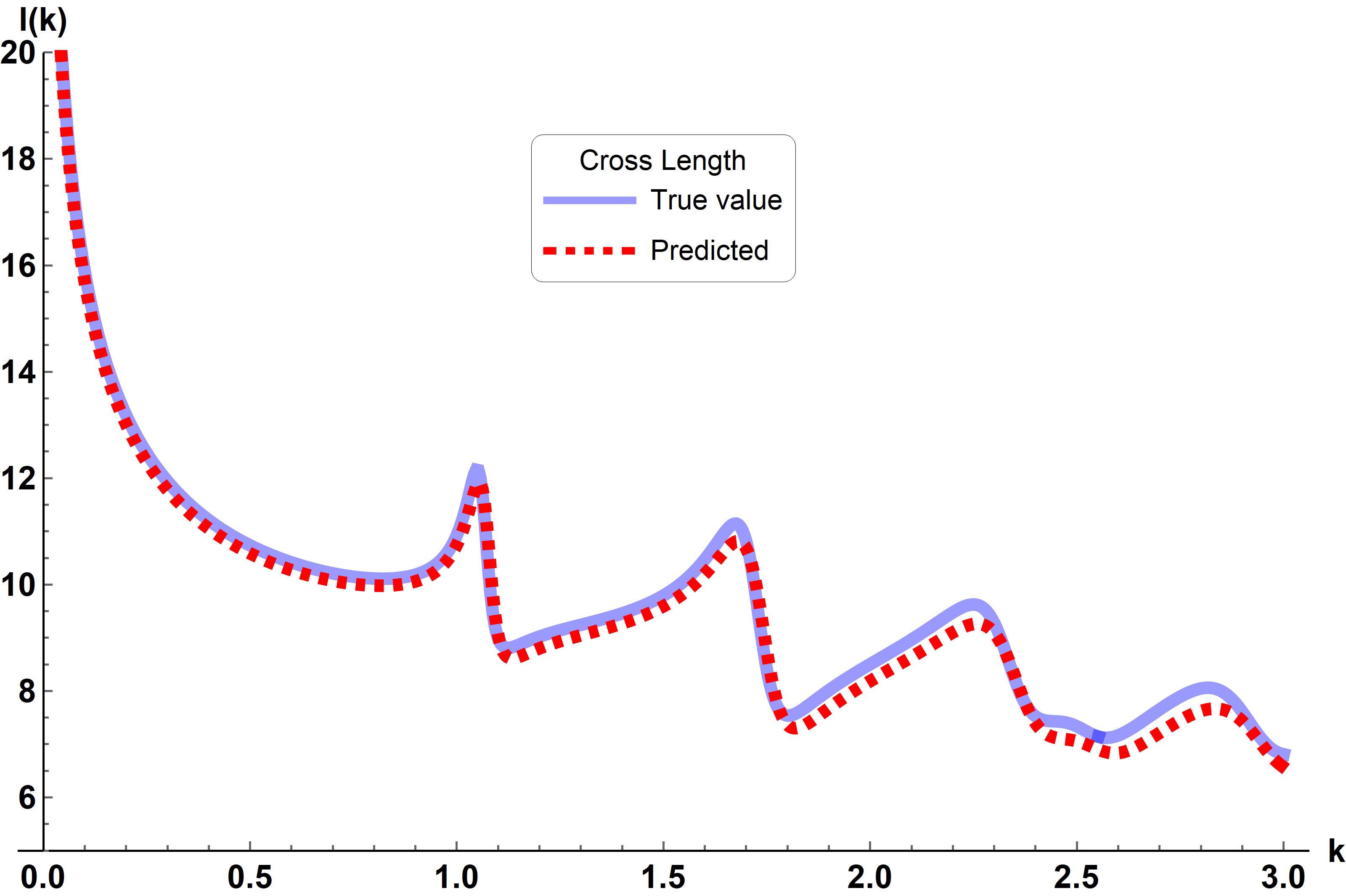}
    \caption{Plot of the scattering cross length. The blue (gray) full line is related to the true values $R=2$ and $\gamma=2$, and the red (black) dashed line to the ``predicted" values $\gamma \approx 1.92$ and $R\approx 1.98$ obtained via the trained neural network. }
    \label{fig: comparacao}
\end{figure}

\begin{figure*}
    \centering
    \includegraphics[width=.48\columnwidth]{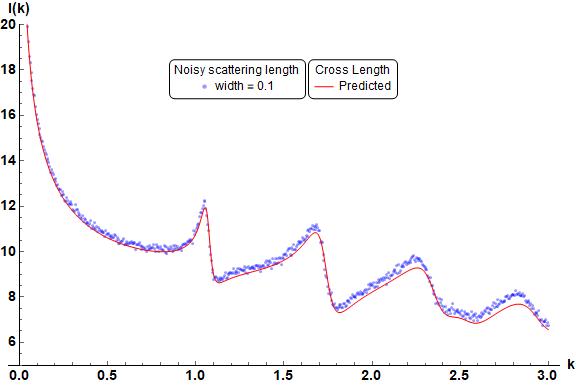}
    \includegraphics[width=.49\columnwidth]{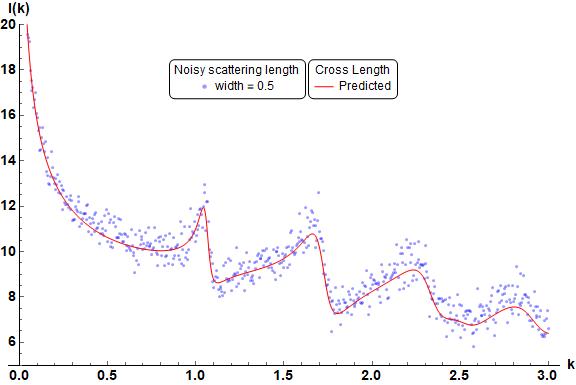}
    \includegraphics[width=.48\columnwidth]{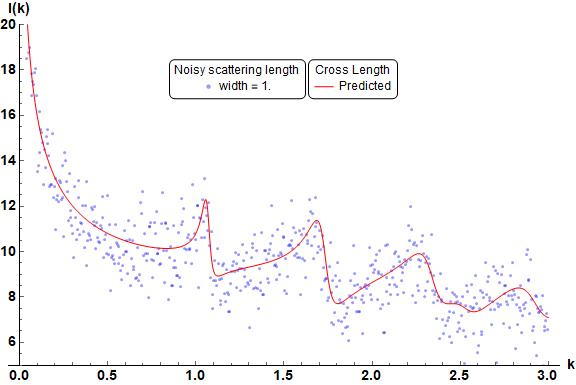}
    \includegraphics[width=.48\columnwidth]{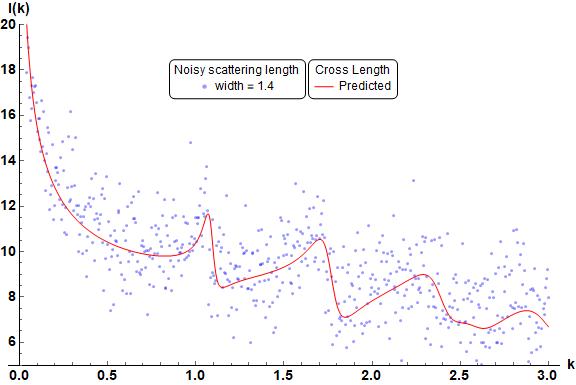}
    \caption{Scatter plot of the noisy scattering cross length with noise width $w=0.1$ (upper left), $w=0.5$ (upper right), $w=1.0$ (bottom left), $w=1.4$ (bottom right). The red full line corresponds to the scattering cross-length calculated with the predicted parameters obtained via the trained neural network.}
    \label{fig: ilustraruido}
\end{figure*}

For central potentials, it is useful to apply the partial wave analysis,
\begin{equation}\label{eq: amplitudePW}
    f(\theta)=\sum_{l=-\infty}^\infty f_n e^{il\theta},
\end{equation}
where
\begin{equation}\label{eq: amplitudefaseshiftPW}
    f_n=\sqrt{\frac{2}{\pi}} \; e^{i\pi/4} \; \sqrt{\frac{1}{k}} \; e^{i\delta_n} \sin{\delta_n}, 
\end{equation}
and $\delta_n$ is the phase shift. One can find an analytic expression for the phase shift after combining eq. (\ref{eq: amplitudemeu}), (\ref{eq: amplitudePW}) and (\ref{eq: amplitudefaseshiftPW}) 
\begin{equation}
    \delta_n= \frac{\log(1+2u_n)}{2i},
\end{equation}
and a relation for the scattering cross length 
\begin{equation}\label{eq: crosslength}
    l(k)=\frac{4}{k}\sum_{n=-\infty}^\infty \sin^2(\delta_n)=-\frac{4}{k}\sum_{n=-\infty}^\infty {\rm Re}\left[u_n\right],
\end{equation}
where ${\rm Re}\left[u_n\right]$ stands for the Real part of $u_n$.
For a chosen $\gamma$ and $R$ it is computed $l(k)$ for several values of $k$. It begins with $k_{min}=0.02$ and ends at $k_{max}=3$ with increments $\Delta k=0.005$, and it is used natural units $m=\hbar=1$. The series of eq. (\ref{eq: crosslength}) was truncated at $n_{max}=20$
\begin{equation}
    l(k)=-\frac{4}{k}\sum_{n=-20}^{20} {\rm Re}\left[u_n\right].
\end{equation}
So, one synthetic data is the group of $603$ values $\mathbf{x}=(m,\hbar,k_{min},k_{max},\Delta k,n_{max},l(k_{min}),...,l(k_{max}))^T$. Those values are organized as a column vector $\mathbf{x}$ and are the input of the neural network. Therefore, we generate $55100$ synthetic data, for different values of $R$ and $\gamma$, where $R$ spams from $0.1$ to $2$ with steps of $0.01$, and $\gamma$ from $0.1$ to $3$ with increment $0.01$.

\subsection{Third Step: Build a neural network}\label{subsec: buildnetwork}
Choosing a specific Neural Network to implement a regression problem is decisive due to the antagonism between the computational time to execute the program and the spend personal time desired to obtain the solution. Among several types of Neural Networks (such as Recurrent Neural Networks, Modular Neural Networks, Convolutional Neural Networks, and more), we choose a Multilayer Perceptron because it has a simple setup and provides excellent results. The number of hidden layers in this work ($15$) is justified at the subsection \ref{subsec: trainednetwork}. Usually, the more hidden layer in the network better is the results, until it starts to overfit. However, for hidden neurons, one can employ the rules\cite{heaton2015artificial}:
\begin{itemize}
    \item The number of hidden neurons should be between the size of the input layer and the size of the output layer.
    \item The number of hidden neurons should be 2/3 the size of the input layer, plus the size of the output layer.
    \item The number of hidden neurons should be less than twice the size of the input layer.
\end{itemize}
 The chosen number in this example was the size of the input plus one-third of it ($804$), and the activation function was the logistic sigmoid. 

\subsection{Fourth Step: The Training}\label{subsec: trainednetwork}
To train the neural network, the synthetic data were randomly separated among three groups, namely the training set, validation set, and test set. The test set has $20\%$ of the total number of synthetic data. The remaining ($80\%$) was allocated between the training and validation sets. $30\%$ of it for the validation set and $70\%$ to the training set. This separation is important to check the accuracy of the network. The error (loss or cost) function $J$ employed is the mean squared difference
\begin{equation}
    J(\mathbf{y},\mathbf{y}')=\frac{1}{N}\sum_{j=1}^N (y_j-y'_j)^2,
\end{equation}
where $\mathbf{y}=(y_1,...,y_N)^T$ is the network output, $N=2$ is the size of the output and $\mathbf{y}'=(y_1',...,y_N')^T$ is the desired output, in other words, the $\gamma$ and $R$ used to produce $\mathbf{x}$. The training method is the stochastic gradient descent with a batch size of $100$ examples, where is important to apply an adaptive learning rate that is invariant to diagonal rescaling of the gradients \cite{kingma2017adam}. However, one should avoid training the neural network directly, because of the vanishing gradient problem. This leads to a network with high bias. 

It is known, that a cascade-correlation learning architecture \cite{cascade} solves this problem. The procedure consists in training the network several times, first with only one hidden layer. Then, one adds another hidden layer and keeps the weights learned previously. At each training, one must check the convergence of the error over the test set and the validation set. If the error calculated over the validation set increases (over each iteration at one training), then you have overfitting. To solve this problem decrease the number of hidden neurons. Finally, it is imperative to apply the network over the test set at the end of each training, because one can visualize the error decreasing until reaching the desired value. In this work, we stop at $15$ hidden layers and obtain an error over the test set of $\sim 10^{-2}$. One can go further (more hidden layers) but this is enough for the purpose of this work.

 After checking the convergence of the parameters, we repeat the training with all the synthetic data. As an example, in Fig. \ref{fig: comparacao} is plotted the scattering cross length calculated considering $\gamma=2$ and $R=2$ (blue full line). Then, it is provided to the neural network as an input, and it ``predicts" the values $\gamma \approx 1.92$ and $R\approx 1.98$. Consequently, is plotted the scattering cross length computed with those values (red dashed line). We calculate the percentage relative difference
 \begin{equation}
     \frac{|\gamma-p_\gamma|}{\gamma} \approx 4.1 \% ,\quad \frac{|R-p_R|}{R} \approx 1.2 \%,
 \end{equation}
where $p_\gamma$ and $p_R$ stands for the ``predicted" values obtained by the neural network. 


\section{Prediction with noise}\label{sec: noise}
The trained neural network can predict accurate values of parameters when the input data has noise. It generated synthetic data $l(k)$ and added Gaussian white noise with different widths. Therefore, it was plotted (Fig. \ref{fig: ilustraruido}) the noisy scattering cross length with its respective prediction to elucidate the procedure. The four plots correspond to the same scattering cross length (same as presented in Fig.\ref{fig: comparacao}), although their difference is the noise width. Along these lines, each example from Fig. \ref{fig: ilustraruido} has a correct prediction for the potential parameters. Here we consider a correct prediction as a percentage relative difference less than $10\%$ for all the parameters. Then, it was generated one thousand examples for each width of the noise, where the parameters were randomly selected between the interval $R \in \left[ 0.1,2\right]$ and $\gamma \in \left[ 0.1,3\right]$. In Fig. \ref{fig: ilustraruido} is plotted the percentage of correct predictions for each noise width $w$. It is shown a decrease in accurate predictions as the value of the noise increase.


\begin{figure}
    \centering
    \includegraphics[width=.7\columnwidth]{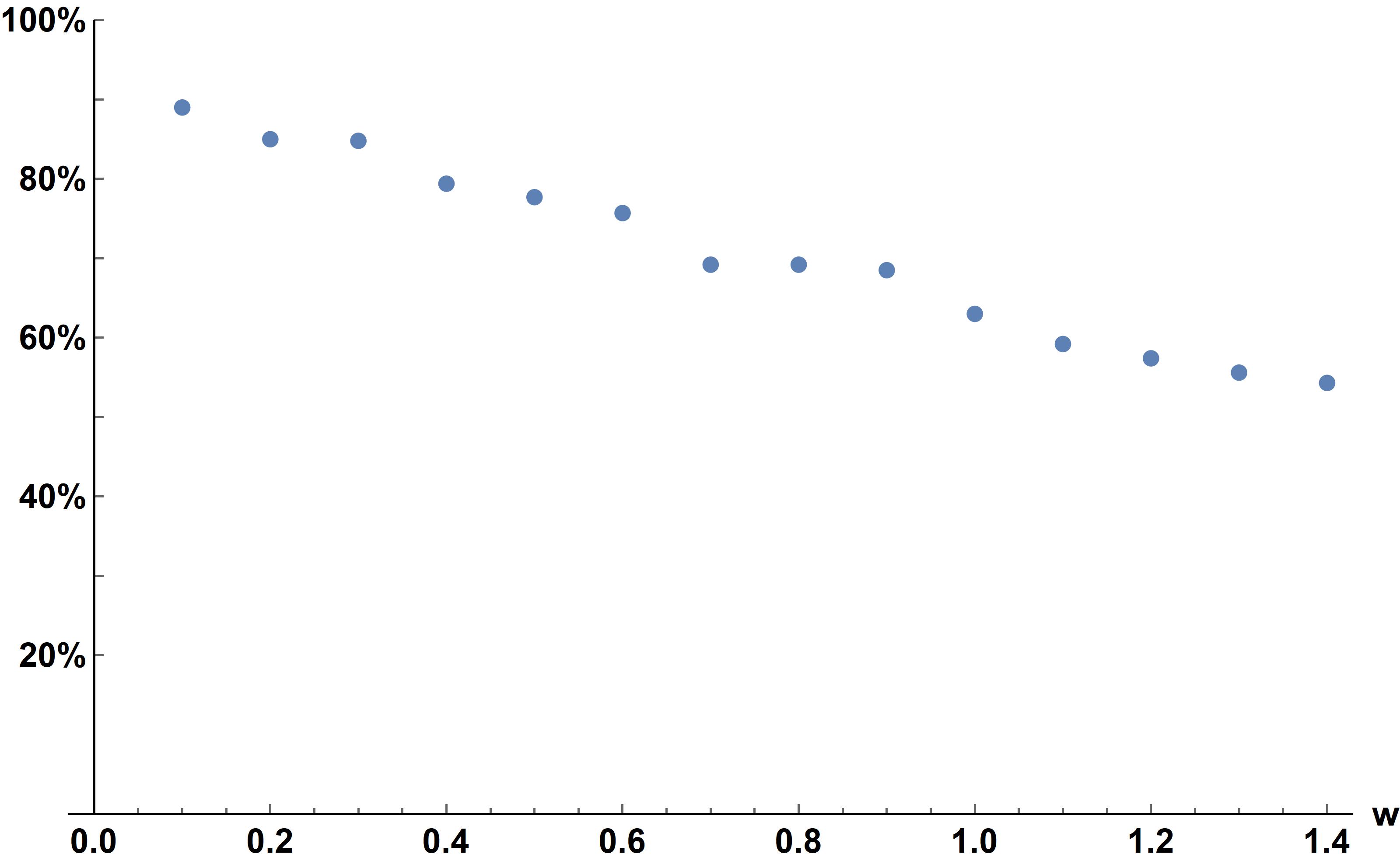}
    \caption{Percentage of correct predictions for each noise width $w$. It is considered as a correct prediction for any example with a percentage relative difference less than $10\%$ for both parameters simultaneously.}
    \label{fig: acertos}
\end{figure}
\section{Conclusion}\label{sec: conclusion}
In this work, we have shown how a simple neural network can predict correct values for potential parameters. We choose a circular boundary-wall potential due to the existence of the analytic solution for the wave function and the scattering cross length. However, the vast majority of potential does not have an analytic solution for the wavefunction nor the scattering cross length (or scattering cross section in 3D problems). Consequently one can obtain it via numeric (boundary integral methods) or approximate (Born approximation) methods. The neural network is able to determine the parameters even with noisy input.


\bibliography{redeNN}

\end{document}